\documentclass{aa}

\usepackage{graphics}
\usepackage{astron}

\newcommand{\sM}{\sigma_{M}}
\newcommand{\sU}{\sigma_{U}}
\newcommand{\sV}{\sigma_{V}}
\newcommand{\sW}{\sigma_{W}}

\newcommand{\cA}{{\cal A}}
\newcommand{\cB}{{\cal B}}

\begin{document}

\title{The LMC distance modulus from Hipparcos RR Lyrae and 
       classical Cepheid data}

\titlerunning{The LMC distance modulus from Hipparcos data}

\author{
   X. Luri\inst{1,2}
   \and A.E. G\'omez\inst{1}
   \and J. Torra\inst{2}
   \and F. Figueras\inst{2}
   \and M.O. Mennessier\inst{3}
}

\thesaurus{}

\offprints{X. Luri}
\mail{xluri@mizar.am.ub.es}

\institute{
           Observatoire de Paris-Meudon,
           D.A.S.G.A.L., URA CNRS 335, 
           F92195  Meudon CEDEX, France
       \and
           Departament d'Astronomia i Meteorologia,
           Universitat de Barcelona,
           Avda. Diagonal 647,
           E08028, Barcelona, Spain
       \and
           Universit\'e Montpellier II,
           GRAAL, URA CNRS 1368,
           F34095 Montpellier CEDEX 5, France
          }

\date{Received ; Accepted }

\maketitle

\begin{abstract}

The LM method \cite{luri96}, designed to exploit the Hipparcos 
data to obtain luminosity calibrations, is applied to derive 
luminosity calibrations for RR Lyrae and classical Cepheids. From these 
calibrations the distance to the Large Magellanic Cloud (LMC) is 
estimated. The distance moduli provided by the two calibrations 
are in good agreement, giving a value of $\sim 18.3^m$, while several
previous calibrations using Hipparcos data provided inconsistent results 
between both types of stars. This result suggest that the Hubble constant 
should have a value of $H_0 \sim 79$ km s$^{-1}$Mpc$^{-1}$.
   
\end{abstract}

\section{Introduction}

The calibration of the absolute magnitudes of RR Lyrae and the classical 
Cepheids is the first step in the determination of the extragalactic distance 
scale, and the recently released Hipparcos data \cite{ESA} allow, for the
first time, its determination on the basis of trigonometric parallaxes.
However, in spite of the high accuracy of these data, few of these stars 
have precise trigonometric parallax measurements: only 12 RR Lyrae and 6 
classical Cepheids have relative errors in trigonometric parallax smaller 
than 30\%.  Due to this limitation, and to the intrinsic difficulty of 
determining distances and absolute magnitudes from trigonometric parallaxes
(several biases may arise from the effects of the observational 
errors and sample censorship, see Brown et al. \cite*{brown}), a careful 
statistical treatment of the data is required to obtain reliable calibrations. 

The difficulty of these estimations is illustrated by the wide range
of values for the distance modulus of the Large Magellanic Cloud (LMC)
obtained from published luminosity calibrations using Hipparcos data: 
from RR Lyrae $18.31^m$ \cite{fernley} (direct determination), $18.63^m$ 
\cite{gratton}, $18.65^m$ \cite{reid} (indirect determinations obtained 
from subdwarf-sequence fitting on globular clusters) and from the classical 
Cepheids $18.44-18.57^m$ \cite{madore}, $18.72^m$ \cite{paturel}, 
$18.70^m$ \cite{feast}.

In this paper luminosity calibrations for both RR Lyrae and classical Cepheids
are obtained using the LM method applied to Hipparcos data. The
results provide compatible values for the LMC distance modulus.

\section{The LM method}

The LM method \cite{luri96} is based on the Maximum-Likelihood estimation.
It includes a detailed model of the luminosity, kinematics and spatial 
distribution of the sample and takes into account its observational censorship 
and observational errors, thus providing estimations free of biases due
to these two factors \cite{luri97}. The interstellar absorption is taken into 
account by using the Arenou et al.  \cite*{arenou} 3D model.

Using the LM method, the parameters of the model used are estimated. The
estimation uses all the available information for the stars in the sample: 
apparent magnitude, galactic coordinates, trigonometric parallax, proper 
motions, radial velocity and any other relevant parameter such as metallicity 
or period. The use of all the observational data is specially important
in the present case because parallaxes alone would not provide a precise
enough calibration (their relative errors being high, even with the Hipparcos
high-precision astrometry). Furthermore, as the estimation is done by 
Maximum-Likelihood, the information given by these observational data is 
included through the {\em Probability Density Function} (PDF) defined by 
the model and the observational errors. Consequently, each individual piece 
of data has its own ``intrinsic weight'' in the solution \footnote{the relative
contribution of parallaxes and proper motions to our solutions will be assessed
in future papers}  and there is no need, as in other methods used for absolute 
magnitude calibration, for any external system to weight the contribution of, 
say, parallaxes or proper motions in the estimation.

\section{RR Lyrae} \label{sec:RR}

The LM method was adapted to determine a mean absolute magnitude and
the corresponding dispersion $\sM$ for the RR Lyrae stars. The distribution 
of metallicities was modeled (and fitted) using normal distributions.
To model the kinematics of the sample a velocity ellipsoid with
means $(U_0,V_0,W_0)$ and dispersions $(\sU,\sV,\sW)$ was adopted.
An exponential galactic disk with scale height $Z_0$ was used to 
describe the spatial distribution. 

On the other hand, the apparent magnitude selection of the sample was also 
taken into account. The Hipparcos catalogue was designed to be complete up 
to an apparent magnitude varying on galactic latitude and spectral type, 
and for fainter magnitudes very heterogeneous selection criteria were used. 
In the case of RR-Lyrae the criteria used to complete the catalogue up to
the Hipparcos magnitude limit is described in Mennessier \& Baglin 
\cite*{menes} and, furthermore, six previously unknown RR-Lyrae were found. 
This observational censorship was modeled in the LM method by assuming the 
sample to be complete up to an apparent magnitude $V_c$ (determined at the 
same time than the rest of the parameters) and with a linear decrease in 
completeness up to the apparent magnitude limit, reflecting the fact that 
fainter RR-Lyrae have a smaller probability to be included.\\

The data used for the RR Lyrae calibration comes from two sources:
astrometric data from the Hipparcos Catalogue \cite{ESA} and intensity-mean 
V apparent magnitudes (calculated from the Hipparcos data), metallicities 
and radial velocities from the compilation of Fernley et al. \cite*{fernley}. 
There are 186 RR Lyrae stars in the Hipparcos catalogue, 6 of them newly 
discovered.  The Fernley compilation contains 144 stars (125 RRab and 19
RRc) reliably classified as RR-Lyrae, which constitute our sample. 

The LM method identified two main groups, constituting the 91\% of the
sample. The first group corresponds to the Halo population and the second 
to the Disk population. The mean magnitudes and metallicities for these 
groups are listed in Table \ref{tab:resRR}.

\begin{table}[h]
  \caption{\em Mean absolute magnitudes and metallicities for 
               RR-Lyrae}
  \label{tab:resRR}
  \begin{center}
  \begin{tabular}{lccc}
           & $\overline{<M_v>}$ & $\overline{[Fe/H]}$ & \% of the sample   \\
   \hline
     Halo  & $0.65 \pm 0.23$ & $-1.51 \pm 0.06$ & $78.3 \pm 2.4$ \\
     Disk  & $0.13 \pm 0.49$ & $-0.45 \pm 0.07$ & $12.7 \pm 1.6$ \\
 \end{tabular}
 \end{center}
\end{table}

Our results can be compared with those reported by Fernley et al. 
\cite*{fernley}. They obtain an estimation of the Halo RR-Lyrae luminosities
from two different methods. After averaging them they adopt a value 
of $<M_v>= 0.77 \pm 0.15$ at $[Fe/H]=-1.53$ is adopted.  The differences 
with our results can be accounted for by the 
different criteria used to separate Halo and Disk. While Fernley  et 
al. \cite*{fernley} use an a priori metallicity criterion  to divide the 
sample into Halo and Disk, our separation is part of the fit, taking 
simultaneously into account the luminosity, the kinematics and the 
metallicity of the stars.\\

Other recent estimates for the halo RR Lyrae luminosities using Hipparcos data 
are inconsistent with ours \cite{reid,gratton}, giving brighter mean absolute 
magnitudes. However, they are indirect estimates based on 
determinations of the subdwarf sequence and they include a posteriori 
corrections of parallax biases that can degrade their precision \cite{brown}.

\section{Classical Cepheids}

For these stars we consider a period-luminosity (PL) relation
(Eq. \ref{eq:relCep}):

\begin{equation} \label{eq:relCep}
  <M_v> = \cA + \cB \: \log (P) 
\end{equation}

It was assumed that for each value of the period the individual values 
of $<M_v>$ are distributed normally around the PL relation with a dispersion 
$\sM$. The periods were modeled (and fitted) using normal distributions. 
The kinematics, spatial distribution and apparent magnitude selection 
were modeled as explained in Section \ref{sec:RR}. The values of 
Oort's constants and the Sun's galactocentric distance were not determined
but adopted to be $A= 14.4$, $B=-12.8$ $\:km\:s^{-1}kpc^{-1}$ and 
$R_\odot=8.5 \: kpc$.

The sample was formed by selecting the classical Ce\-pheids ($\delta$-Cepheids) 
of the Hipparcos catalogue \cite{ESA}. The known sinusoidal $\delta$-Cepheids
(overtone Cepheids) were eliminated. All data (including periods) were 
taken from the Hipparcos catalogue except the radial velocities, taken 
from the Hipparcos Input Catalogue \cite{turon}. The arithmetic-mean apparent
magnitudes given by Hipparcos were compared with the intensity-mean
apparent magnitudes given in the {\it David Dunlap Observatory Database 
of Galactic Classical Cepheids} 
\footnote{http://ddo.astro.utoronto.ca/cepheids.html} and no systematic 
difference was found (mean difference $0.01^m \pm 0.01$). Thus, the
Hipparcos data were preferred due to their higher homogeneity. The final
sample contains 219 stars.\\

\noindent Two determinations of the PL relation were obtained:

\begin{description}

 \item[\bf Relation 1:] following the approach taken by Feast \& Catchpole 
      \cite*{feast}, the PL slope ($\cB$) was fixed to the value for the 
      LMC, $\cB=-2.81$ \cite{caldwell}. The underlying hypothesis is 
      that the slope of the PL relation is (except for a small metallicity 
      correction) universal, so the slope for the LMC Cepheids can be used
      and only the zero point of the relation remains to be determined.

 \item[\bf Relation 2:] both the slope and the zero point are determined.

\end{description}

\noindent In both cases the LM method identified a small secondary group, 
but the most part of the sample (91\%) belongs to the 
main group. The two solutions obtained for the PL relation of this main 
group are presented in Table \ref{tab:resCep}. 

\begin{table}[htb]
  \caption{\em Period-luminosity relations for the classical Cepheids}
  \label{tab:resCep}
  \begin{center}

  \begin{tabular}{c}
      Relation 1  \\
   \hline
       \\
      $<M_V> = -2.81 \log (P) - (1.05 \pm 0.17)$\\
  \end{tabular}

  \vspace{0.5cm}

  \begin{tabular}{c}
      Relation 2  \\
   \hline
       \\
      $<M_V> = -2.12 \log (P) - 1.73$\\
      $\epsilon_{<M_v>} = 0.20 + 0.08 \log(P)$
      
  \end{tabular}

  \end{center}
\end{table}

In the case of Relation 2, the slope and zero point of the PL relation 
were not used as parameters directly determined by the method due to the high
correlation between them could degrade the precision of 
the numerical method used to maximize the likelihood. Instead, two points
of the PL relation (at two arbitrary values of the period) were 
determined (thus defining the linear relationship) and the slope and
zero point were calculated from them. Consequently, the errors in the 
estimates of the zero point and the slope cannot be given independently 
and, instead, an estimation of the expected error in the absolute magnitude 
($\epsilon_{<M_v>}$) is given as a function of $\log(P)$.

For the Cepheids, unlike the RR Lyrae, the errors in the interstellar 
absorption from the Arenou et al. \cite*{arenou} model (hereafter AGG) can
be high (most of the stars are located in the galactic plane
and at higher distances than the RR Lyrae). To obtain the value of the
intrinsic dispersion we should take into account that the value of the 
magnitude dispersion given by the LM method is the result of this
dispersion $\sigma_M$ and the errors in the estimation of the interstellar 
absorption $\sigma_{A_v}$:
$\sigma^2_{M \: total}= \sigma^2_M + \sigma^2_{A_v}=(0.8^m \pm 0.1)^2$.

The AGG model provides estimations of the errors in the values of the 
interstellar absorption. Using these estimations to correct the total 
dispersion, the value of the dispersion of the sample around the PL 
relation can be estimated as $\sM=0.4^m \pm 0.2$. In any case, the PL 
relations obtained do not depend on the value of this parameter, as shown 
by Monte-Carlo simulations. 

A recent result for the PL relation from Hipparcos data 
is the one of Feast \& Catchpole \cite*{feast} (hereafter FC):

\begin{flushleft}
  $<M_v> = -2.81  \log (P) + (-1.43 \pm 0.10)$
\end{flushleft}

This result can be compared with our Relation 1 (both rely on the hypothesis 
of a known slope $\cB=-2.81$). Our zero point is $0.38^m$ fainter than that 
given by FC but, before a discussion of this difference some details about 
the FC approach are necessary. To determine the zero point of the PL relation 
FC use the following method. Given Eq. \ref{eq:relCep} and Pogson's law, the 
following relation holds:

\begin{equation} \label{ef:relFC}
  10\: ^{0.2 \cA} = 0.01 \pi \: 10\: ^{0.2 \: [<V>_0 - \cB \log (P) ]},
\end{equation}

\noindent where $<V>_0$ is the intrinsic apparent magnitude, i.e.
corrected for  interstellar absorption. For each star the quantity
$Q=10\: ^{0.2 \cA}$ can be estimated and the zero point of the PL
relation $\cA$ calculated from the mean value obtained for all the
stars.

This method of estimating the zero point of the PL relation has the
advantage of avoiding the direct calculation of absolute magnitudes 
from parallaxes, which can lead to a bias (even when using Hipparcos 
unbiased parallaxes) if not treated properly \cite{brown}. Instead, 
the parallaxes are directly averaged and the zero point estimated 
from the average, minimizing this source of bias. However, the method is 
highly sensitive to any error in the exponent of the right hand side of 
Eq. \ref{ef:relFC}, including any effects on the magnitude distribution 
(like Malmquist bias) or the reddening correction. On the other hand, the 
weighting system used by FC to obtain the mean value of $Q$ for the sample 
can have some undesired side-effects: as the weigth of each star is 
proportional to $\frac{1}{Q_i^2}$, being $Q_i$ the individual value of 
$Q$ for the star, stars with low walues of $Q_i$ are favoured in the final 
solution. Furthermore, due to the weighting only a (relatively small) 
fraction of the sample significantly contributes to the solution, so
arising the issue of how representative of the whole population are these 
contributing stars. 

The impact of these effects on FC method is difficult to evaluate, but
Monte-Carlo simulations of realistic samples show that the zero point
given by FC could be slightly (about $0.05-0.1^m$) too bright due to them,
contributing to explain in part the difference with our results.\\

On the other hand, when the LM method is applied using the absorption 
correction method given in FC instead of using the Arenou et al. 
\cite*{arenou} model, a value of $\sigma_{M \: total}= 0.7^m \pm 0.2$ 
is obtained. This result suggests that the combination of 
$\sM$ and errors in the absorption estimation $\sigma_{A_v \: FC}$ 
gives a total dispersion higher than the estimated by FC. The
PL relation, $<M_V> = -2.04 \log (P) - 1.74$, and the kinematics and 
scale height obtained do not differ significantly from the 
results in Relation 2.\\

Our second relation (Table \ref{tab:resCep}) gives a slope of the PL 
relation less steep than the one given by Caldwell \& Laney \cite*{caldwell}, 
but consistent with the results of Szabados \cite*{szabados} from Hipparcos data 
for nine non-binary Cepheids with short periods. 

Further analysis to determine the slope of the PL relation are being carried on 
using the LM method and the preliminary results suggest a different 
behavior in the short and long period regions, possibly due to the effects of 
undetected overtone cepheids in the short period region.

\section{The LMC distance modulus}

The calibrations presented in this paper were used to determine the
mean distance modulus of the LMC. The results are presented
in Table \ref{tab:LMC} and they were obtained as follows:

\begin{description}

  \item[RR Lyrae:] to calculate the distance modulus of the LMC using
                   RR-Lyrae data, a value of the slope of the metallicity-
                   luminosity relation is needed. Although the 
                   value of this slope could be determined using the 
                   LM method, an adopted value was used here,
                   leaving for a forthcoming longer paper the 
                   discussion of this parameter. Notice,
                   however, that the mean magnitude determined here
                   corresponds to a value of metallicity ($-1.51$)
                   close to the mean value of the LMC RR-Lyrae ($-1.8$)
                   so the resulting distance modulus does not depend
                   strongly on the value of the slope adopted.

                   Following the approach of Fernley et al. \cite*{fernley}
                   a slope of $0.18$ was adopted. Using this value and the
                   results for the Halo RR-Lyrae given in Table 
                   \ref{tab:resRR}, a metallicity-luminosity relation was
                   obtained and applied to the RR Lyrae data given in Walker 
                   \cite*{walker} (individual reddening estimates used).

  \item[Cepheids (FC revised):] the FC estimation of the LMC distance
                   modulus was changed by $0.38^m$ to reflect
                   the change in zero point in our Relation 1.

  \item[Cepheids (Rel. 1 \& 2):]  the PL relations given in Table
                   \ref{tab:resCep} were applied to the Cepheid
                   data given in Paturel et al. \cite*{paturel};
                   a mean reddening correction of $E_{B-V}=0.1$
                   \cite{freedmanA} and a metallicity correction
                   of $+0.042$ \cite{laney} were applied.

\end{description}

\begin{table}[h]
  \caption{\em Distance modulus of the LMC using this paper's luminosity
               calibrations}
  \label{tab:LMC}
  \begin{center}
  \begin{tabular}{ll}
   \hline
       & $m_o-M$   \\
   \hline
     RR Lyrae                & $18.37 \pm 0.23$ \\
     Cepheids (FC revised)   & $18.32 \pm 0.17$ \\
     Cepheids (Rel. 1)       & $18.29 \pm 0.17$ \\
     Cepheids (Rel. 2)       & $18.21 \pm 0.20$ \\
 \end{tabular}
 \end{center}
\end{table}

The results of this paper reconcile the distance modulus estimations 
of the LMC based on RR Lyrae and those based on the classical Cepheids.
Moreover, they are consistent with the upper limit of $18.44 \pm 0.05$
derived by Gould \& Uza \cite*{gould} from the analysis of the
SN 1987A supernova ``light echo''. The adoption of a value of $18.3^m$ 
for the distance modulus implies that the Hubble constant should now
have a value of $H_0=79$ km s$^{-1}$Mpc$^{-1}$, in contrast to the 
value of  $H_0=73$ km s$^{-1}$ given by Freedman et al. \cite*{freedmanB}.

\begin{acknowledgements}
  We thank J. Fernley and R. Garrido for providing us 
  with some of the RR-Lyrae data used in this paper.
  This work was supported by the CICYT under contract
  ESP97-1803 and by the PIC program (CNRS PICS 348, CIRIT).
\end{acknowledgements}


\end{document}